\documentclass{PoS}

\title{Light Neutralinos in the NMSSM}

\ShortTitle{Light Neutralinos in the NMSSM}

\author{\speaker{Daniel Albornoz V\'asquez}\thanks{In collaboration with Genevi\`eve B\'elanger and C\'eline B\oe hm.}\\
        LAPTH, U. de Savoie, CNRS, BP 110, 74941 Annecy-Le-Vieux, France\\
        E-mail: \email{albornoz@lapp.in2p3.fr}}

\abstract{Next-to-Minimal Supersymmetric Standard Model neutralino dark matter candidates in the $(1-15)$ GeV range are found with a Markov Chain Monte Carlo scanning code. A very light, singlet-like Higgs and/or CP-odd Higgs is needed in order to achieve such light masses. Light scalar Higgses can yield large direct detection cross sections, suitable for an account of the recent claims for signals. Neutralino annihilations via resonant light pseudoscalar Higgs exchanges can overproduce the indirect yield of $\gamma$-rays from dwarf spheroidal galaxies. Hence, direct and indirect detection experiments probe different configurations of the NMSSM with a light neutralino DM candidate.}

\FullConference{The 2011 Europhysics Conference on High Energy Physics-HEP 2011,\\
		July 21-27, 2011\\
		Grenoble, Rh\^ne-Alpes France}

\begin{document}
The search for light dark matter (DM) candidates has been very active in the past years after encouraging direct detection results. For example, the CoGeNT collaboration claims to have "an irreducible excess of bulk-like events below 3 keV"~\cite{Aalseth:2010vx}. At that energy range, the excess could be interpreted as elastic scattering events of DM particles with a mass of (6-12) GeV and spin independent DM-nucleon interactions around $10^{-40}$ cm$^2$.

The neutralino ($\chi$) is a DM candidate in the Minimal Supersymmetric Standard Model (MSSM) and Next-to-MSSM (NMSSM), which could fulfil such requirements. The NMSSM is a simple extension of the MSSM, obtained by the inclusion of an SU(2) singlet to the Higgs sector (already containing the two MSSM doublets), thus solving the $\mu$-problem of the MSSM~\cite{Ellwanger:2009dp}. The Higgs spectrum gets enchanced to 3 CP-even (H) and 2 CP-odd neutral (A) bosons. Singlet-like lightest Higgses could be almost decoupled from the SM and escape collider constraints, allowing $M_{H_1} \ll 100$ GeV and/or $M_{A_1} \ll 100$ GeV. The $\chi$ is a 5-state mix of neutral gauginos, neutral higgsinos and the singlino. Particle physics constraints obtained at LEP, Tevatron and other colliders make the task of producing NMSSM $\chi$ lighter than 30 GeV rather intricate. The parameter space is multidimensional, and the need for light $\chi$ mass ($m_{\chi}$) implies fine-tuned configurations. In order to find configurations satisfying all constraints (from electroweak obervables, $B$-meson decays and oscillations, the Higgs sector, masses of unobserved particle, the DM density as measured by WMAP) and providing light $\chi$ DM, a Markov Chain Monte-Carlo code (MCMC) was developed~\cite{Vasquez:2010ru}. The MCMC scans yielded $m_{\chi}$ down to 1 GeV in the NMSSM~\cite{Vasquez:2010ru}, as well as broad interaction ranges~\cite{Vasquez:2011js} interesting for direct and indirect searches\footnote{For another account of neutralinos associated to light Higgs bosons, see~\cite{Draper:2010ew}.}.

\begin{figure}[ht]
\centering		
\includegraphics[scale=0.21,natwidth=6cm,natheight=6cm]{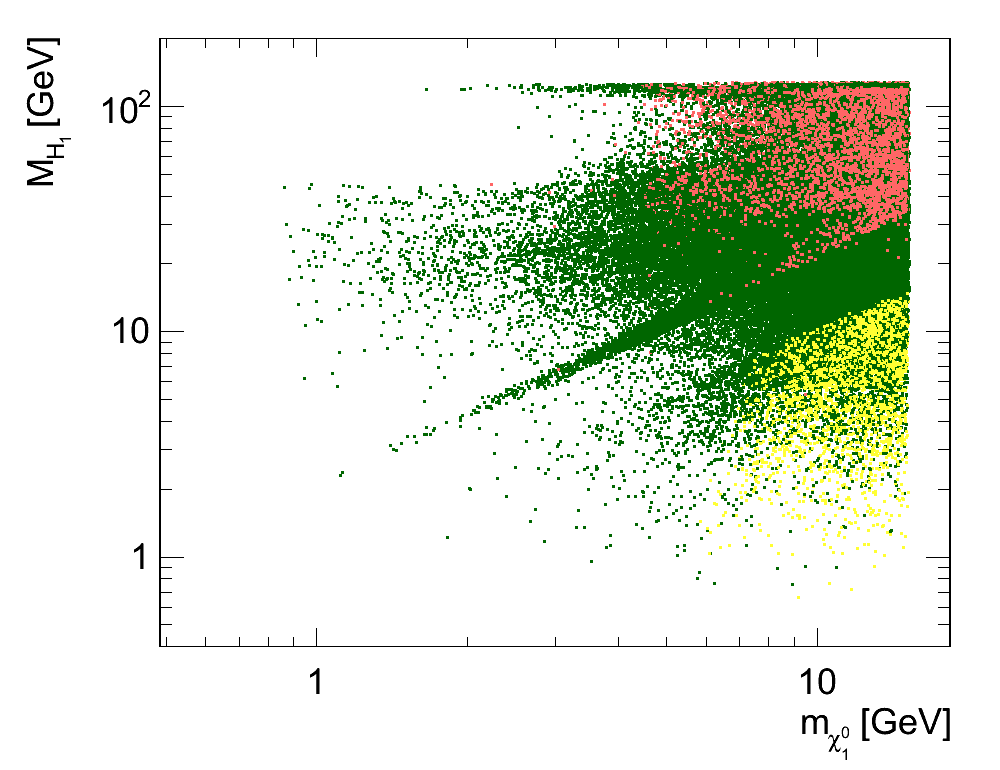}
\includegraphics[scale=0.21,natwidth=6cm,natheight=6cm]{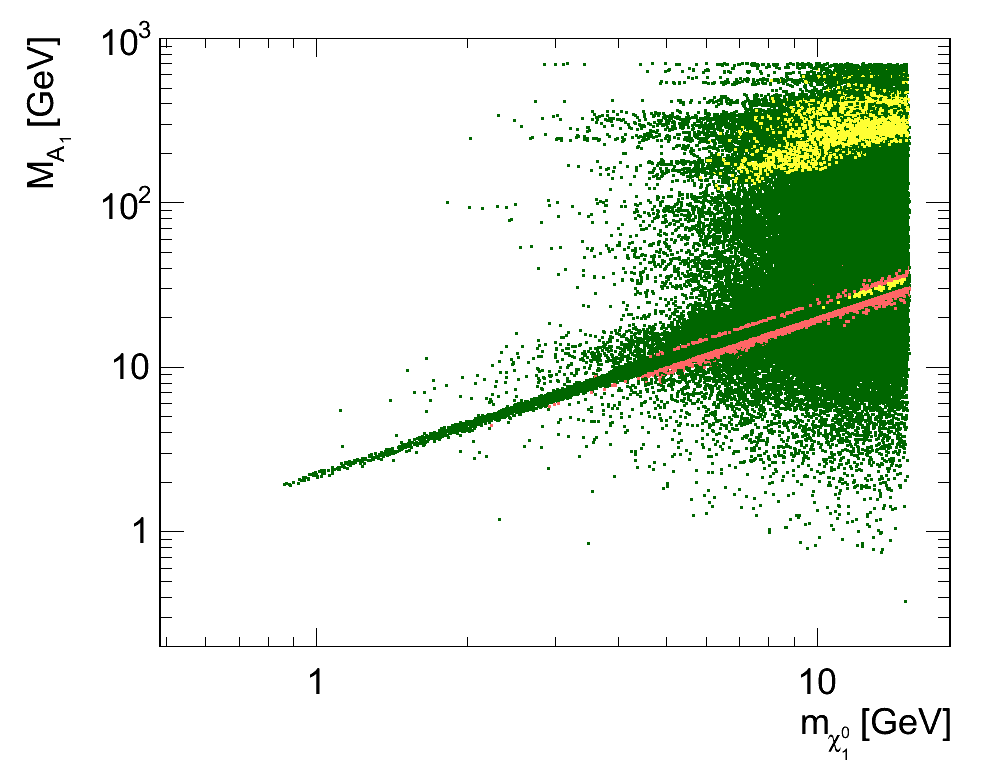}
\caption{Lightest scalar (left) and pseudoscalar (right) Higgs masses as a function of neutralino mass. Red: points excluded by Fermi-LAT; yellow: points above the XENON100 limits; green: safe points.}
\label{fig:higgs}
\end{figure}

\begin{figure}[ht]
\centering		
\includegraphics[scale=0.21,natwidth=6cm,natheight=6cm]{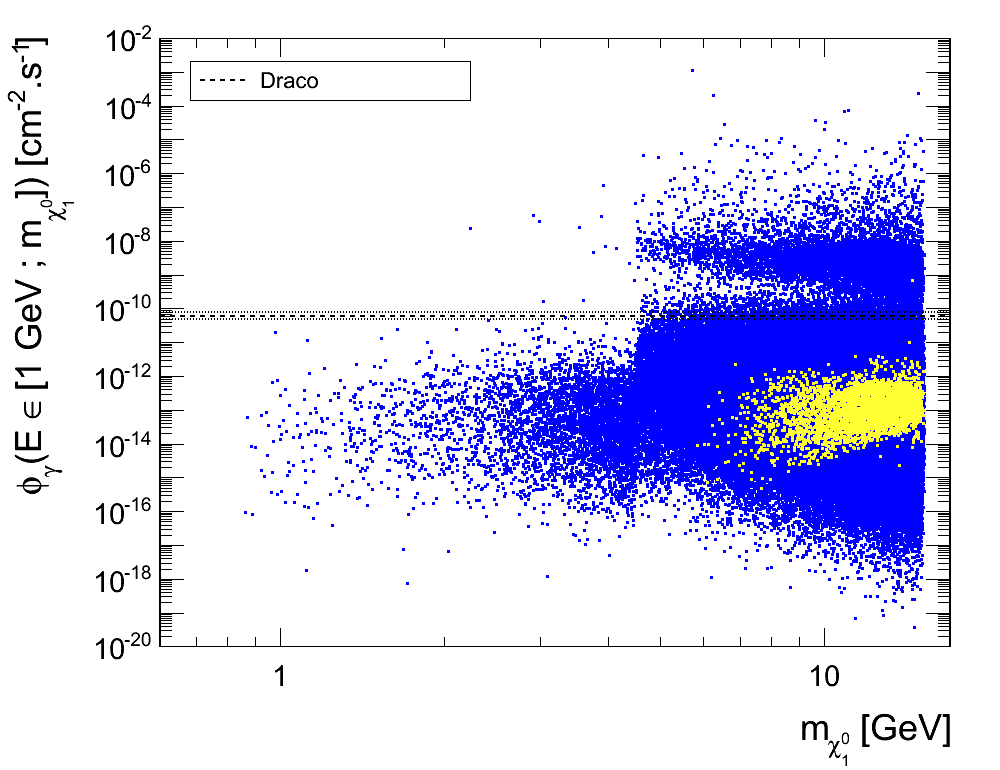}
\includegraphics[scale=0.21,natwidth=6cm,natheight=6cm]{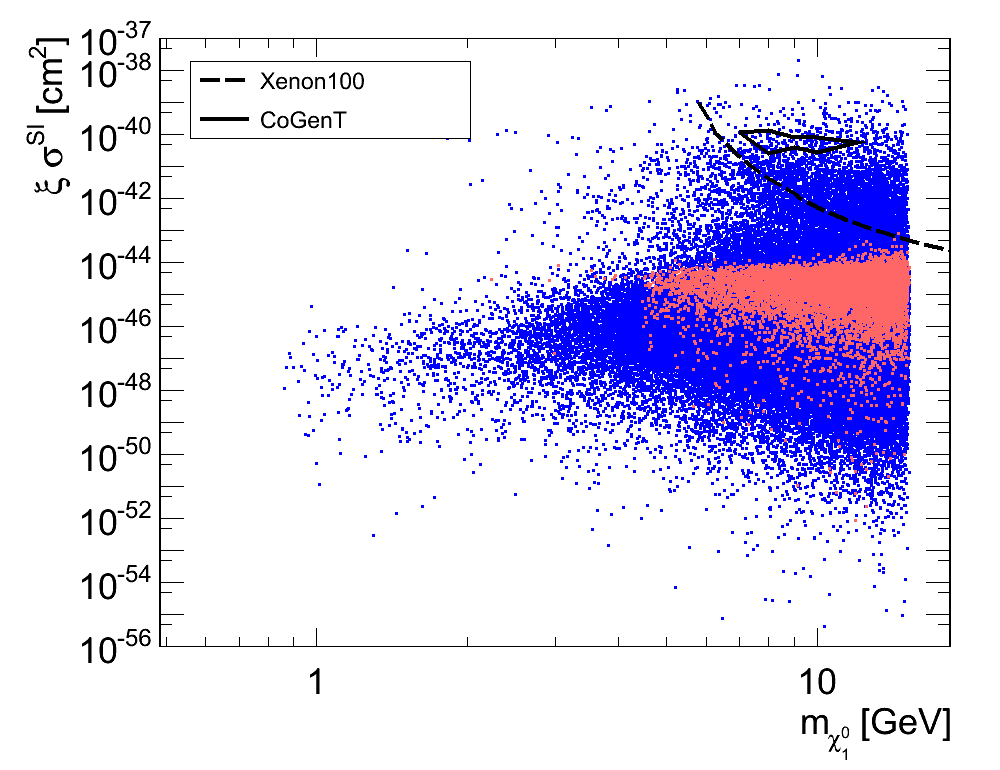}
\caption{Left: $\gamma$-ray flux expected from neutralino annihilations at the Draco dwarf spheroidal galaxy as function of neutralino mass, along with Fermi-LAT limits~\cite{Abdo:2010ex}; yellow: points above the XENON100 limits; blue: all the rest. Right: spin independent elastic scattering cross sections of neutralinos and nucleons as a function of neutralino mass, along with XENON100 limits~\cite{Aprile:2011hi} and the CoGeNT preferred region~\cite{Aalseth:2010vx}; red: points excluded by Fermi-LAT; blue: all the rest.}
\label{fig:astro}
\end{figure}

A resonant exchange of a pseudo-scalar or a scalar Higgs allows to reach large enough annihilation rates, so the lightest $\chi$ gets an acceptable relic density~\cite{Belanger:2005kh}. Fig.~\ref{fig:higgs} shows that for both Higgses the resonance relation $M_{H_1,\;A_1}\simeq2m_{\chi}$ is well represented. Annihilation rates at freeze-out can differ from those in galaxies, where $\chi$ are expected to travel at speeds of $\sim10^{-3}c$: while scalar exchanges are suppressed at low exchanged momentum, resonant pseudoscalar exchanges yield large $\gamma$-ray fluxes. The left panel in Fig.~\ref{fig:astro} shows the $\gamma$-ray flux expected from $\chi$ annihilations in the Draco dwarf spheroidal galaxy, computed using the line-of-sight integral given in~\cite{Abdo:2010ex}. Points overpredicting the flux with respect to the Fermi-LAT limits, are tagged in red in Fig.~\ref{fig:higgs}. One can readily see that those points correspond to resonant pseudocalar exchanges.

Regarding spin independent elastic scattering, $\chi$ and quarks interact mainly via the exchange of a Higgs boson. At low exchanged momentum only the pseudoscalar channel is suppressed. Light scalar exchanges satisfy $\sigma^{SI}\propto M_{H_1}^{-4}$, hence large interactions can occur for the lightmost CP-even Higgses. The right panel in Fig.~\ref{fig:astro} shows these interactions as a function of $m_{\chi}$. Notice that there are configurations falling in the CoGeNT contour. All points above the XENON100 limit are tagged in yellow in Fig.~\ref{fig:higgs}. We can see that all of them correspond, indeed, to light scalar Higgs bosons.

The NMSSM provides light $\chi$ associated to light Higgs bosons. In particular, light scalar Higgs exchanges could explain direct detection signals, while resonant light pseudoscalars could overproduce $\gamma$-rays. These two cases are complementary as it can be seen in Fig.~\ref{fig:astro}: each detection technique constrains one of the Higgs configurations.

\end{document}